\documentstyle[aps,prl,epsfig,floats]{revtex}
\begin{document}
\draft
\twocolumn[
\hsize\textwidth\columnwidth\hsize\csname@twocolumnfalse\endcsname

\title{Phase transition induced hydrodynamic instability
and Langmuir-Blodgett Deposition}
\author{Kok-Kiong Loh, Avadh Saxena, Turab Lookman}
\address{Theoretical Division, Los Alamos National Laboratory,
Los Alamos, NM 87545}
\author{Atul N. Parikh}
\address{Department of Applied Science, University of California at
Davis, Davis, CA 95616}
\date{\today}
\maketitle
\begin{abstract}
We propose a model to understand periodic oscillations relevant to
the origin of  mesoscopic channels formed during a
Langmuir-Blodgett deposition observed in recent experiments \{M.
Gleiche, L.F. Chi, and H. Fuchs, Nature {\bf 403}, 173 (2000)\}.
We numerically study one-dimensional flow of a van der Waals fluid
near its discontinuous liquid-gas transition  and find that
steady-state flow becomes unstable in the vicinity of the phase
transition. Instabilities leading to complex periodic
density-oscillations are demonstrated at some suitably chosen sets
of parameters.
\end{abstract}
\pacs{68.55.-a, 68.18.+p, 68.55.Ln, 68.60.-p}
]

Studies of pattern formation in non-equilibrium systems have
attracted considerable attention due to their relevance in
understanding fundamental physics  and the potential for
applications in a wide range of emerging technologies. The
spontaneously formed patterns can be converted into functional
structures when the pattern forming mechanism is properly
controlled and designed. Typically, this class of pattern
formation involves systems flowing at conditions around which the
properties undergo dramatic changes. The patterns are a result of
the occurrence and propagation of a phase transition or chemical
reaction, the Belousov-Zhabotinsky reaction \cite{Merzhanov} and
capillary flow of molten polymer \cite{JDShore} being two
examples.  One of the key issues to be understood in these systems
is the nontrivial coupling between the system specific `phase
transition' and the hydrodynamics.

Our work has been motivated by recent experiments that have
demonstrated the formation of macroscopic arrays of
submicron-sized channels during a Langmuir-Blodgett deposition
\cite{exp}. Mechanisms based on  stick-slip conditions have been
proposed to describe the phenomenon \cite{exp,stick-slip}.
However, in this Letter, we  propose a different scenario in which
density oscillations occur in the flow of a single-component fluid
at conditions tuned appropriately near its discontinuous
liquid-gas phase transition. The system fails to support
steady-state flow at these conditions due to the absence of a
mechanically stable uniform state in a certain density interval.
The instability leads to periodic oscillations in density or
alternating appearance of liquid and gaseous phases in the system.
The oscillations can be potentially useful in producing large
arrays of periodic structures and we propose this as a possible
mechanism for the experiments \cite{exp} mentioned above. The
problem of fluid flow in the vicinity of a phase transition has
been widely studied \cite{Onuki}. A closely related subject on
propagation of the phase transition front has also been
investigated in detail\cite{OInomoto}. However, we are not aware
of any previous work on unstable one-dimensional flow of a
single-component fluid that leads to periodic density-oscillations
of this nature. A realization of such a system is a fluid flowing
in a narrow channel, or higher dimensional flow in which spatial
variations of the system properties are suppressed in the
direction transverse to the fluid velocity. The flow of surfactant
molecules in a typical Langmuir-Blodgett deposition is an example
in two-dimensions \cite{Petrov,Bruinsma}.

Our analysis begins with the standard hydrodynamics at constant
temperature consisting of the continuity equation and the
Navier-Stokes equation for a one-dimensional flow of compressible
fluid as follows:
\begin{eqnarray}
\frac{\partial \rho}{\partial t} +\frac{\partial \rho v}{\partial x}&=&
0,\label{continuity}\\
\rho\left(\frac{\partial v}{\partial t}+v\frac{\partial
v}{\partial x}\right) &=&-\frac{\partial P}{\partial x}
+\eta\frac{\partial v^2}{\partial x^2} \label{NavierStokes},
\end{eqnarray}
where $\rho$ is the density, $v$ is the velocity and $\eta$ is the
viscosity. The main feature of the system is contained in the
quantity $P$, the local pressure of the inhomogeneous fluid. It is
expressed as \cite{REvans}
\begin{eqnarray}
P = \rho\frac{\delta \Psi[\rho]}{\delta
\rho}-\psi(\rho)+\frac{\kappa}{2}\left( \frac{\partial
\rho}{\partial x}\right)^2,\label{eos}
\end{eqnarray}
which is also the equation of state of the fluid. Here, the
quantity $\Psi[\rho]$ denotes the Helmholtz free energy density
for an inhomogeneous fluid \cite{REvans} given by,
\begin{eqnarray}
\Psi[\rho]=\int\left\{\frac{\kappa}{2}\left(\frac{\partial \rho}
{\partial x}\right)^2+\psi(\rho)\right\}dx,
\end{eqnarray}
where $\psi(\rho)$ is the Helmholtz free energy density of the
homogeneous system and $\kappa$ is a phenomenological constant
associated to the gradient term with a meaning that is specific to
the particular system chosen. For a van der Waals (vdW) fluid, the
Helmholtz free energy density of a uniform system is given by
\begin{eqnarray}
\psi(\rho)=8\rho T\ln\frac{\rho}{3-\rho}-3\rho^2,\label{vdWHfe}
\end{eqnarray}
where $T$ is the temperature and  $\kappa$ is proportional to the
surface tension of the liquid-gas interface. The reason for
choosing a vdW fluid for our investigation is that it is one of
the simplest descriptions that includes a discontinuous liquid-gas
transition and a region in the $P-\rho$ plane where the uniform
fluid is unstable (negative compressibility) below the critical
temperature $T_c\equiv1$. We note that the quantities $P$, $\rho$
and $T$ in Eqs.~(\ref{eos}) and (\ref{vdWHfe}) have been
normalized by their respective values at the critical point for
convenience. We also remark that the fluid is maintained at
thermal equilibrium at all times, that is, Eq.~(\ref{eos}) always
holds. It is the mechanical stability that is being investigated
here.

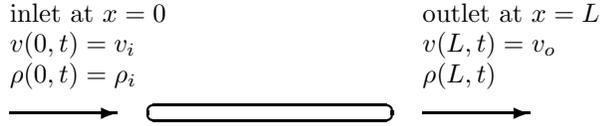
\begin{figure}
\begin{center}
\setlength{\unitlength}{.08in}
\begin{picture}(40,12)(10,15)
\thicklines  \put(38,24){outlet at $x=L$} \put(38,20){$\rho(L,t)$}
\put(38,22){$v(L,t)=v_o$} \put(11,24){inlet at $x=0$}
\put(11,20){$\rho(0,t)=\rho_i$} \put(11,22){$v(0,t)=v_i$}
\put(11,18){\vector(1,0){7}} \put(28,18){\oval(16,1)}
\put(38,18){\vector(1,0){7}}
\end{picture}
\end{center}
\caption{Schematics of the flow configurations. The quantities
$v_i$ and $\rho_i$ are, respectively, the inlet velocity and
density while the outlet velocity is denoted by $v_o$.}
\label{schematics}
\end{figure}

A numerical algorithm has been developed to solve
Eqs.~(\ref{continuity}) and (\ref{NavierStokes}) using the
forward-time-centered-space approach described in
Ref.~\cite{NumRcp}. We will restrict ourselves to a special case
depicted in Fig.~\ref{schematics}:  a flow in the region
$x\in[0,L]$ subject to the following set of boundary conditions
\begin{eqnarray}
v(x<0,\; t)&=&v_0,\label{vinlet}\\
v(x>L,\; t)&=&v_L,\label{voutlet}\\
\rho(x<0,\; t)&=&\rho_0,\label{ninlet}\\
\rho(x>L,\; t)&=&\rho(L,\;t).\label{noutlet}
\end{eqnarray}
The fluid velocity has been arbitrarily chosen so that it flows
from left to right. The positions $x=0$ and $x=L$ correspond to,
respectively, the inlet and the outlet. The quantities $v_0$ and
$v_L$ are, respectively, the velocities at the inlet and the
outlet. The density at the inlet is kept at $\rho_0$. This set of
boundary conditions describes a situation where both the fluid
density and velocity are kept fixed at the inlet, and the fluid
density at the outlet is allowed to vary with time while keeping
the velocity constant. Under this special circumstance, the
instability of the flow is necessarily reflected in the time
dependence of the density at the outlet and we systematically
analyze $\rho(L,\;t)$.  We remark that there are a total of six
parameters involved, namely, $v_0$, $v_L$, $\rho_0$, $T$, $\eta$
and $\kappa$. In general, the solutions to Eqs.~(\ref{continuity})
and (\ref{NavierStokes})~ are complex and possess many degrees of
freedom. Here, we will focus our discussions on the results for a
restricted sets of parameters.

\begin{figure}
\centerline{\epsfig{file=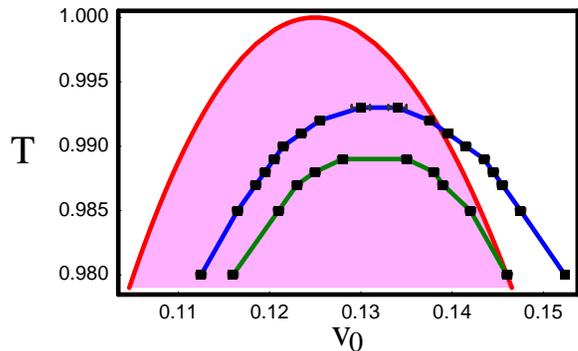, width=3in}} \caption{The
comparison between the computed instability boundary (blue) and
that estimated from the spinodal boundary (red) for $\rho_0=0.8$,
$v_L=0.1$, $\eta=0.01$ and $\kappa=0.0001$. Also shown in green is
the boundary between the simple and complex oscillatory behavior
in the unstable region.} \label{v0pb}
\end{figure}

When the density at the inlet $\rho_0$ and the velocity at the
outlet $v_L$ are kept fixed, the behavior of $\rho(L,\;t)$ as
$v_0$ varies can be understood intuitively as follows:  if one
insists on a steady-state flow, the density at the outlet
$\rho(L,\;t)$ has to be
\begin{eqnarray}
\rho(L,\;t)=\frac{v_0\rho_0}{v_L}.
\end{eqnarray}
Since Eq.~(\ref{eos}), the equation of state of the fluid, allows
for a region where the compressibility is negative (spinodal
region), mechanical instability is thus inevitable if
$v_0\rho_0/v_L$ falls within the spinodal region. A rough estimate
of the stability boundary, assuming the system is uniform, is
given by $v_L$'s for which $\rho(L,\;t)$ coincides with the
spinodal boundary (i.e., state of the fluid where compressibility
diverges). Our numerical results indeed show that the steady state
solution fails to be stable in  some regions of the parameter
space, restricted to the $v_0-T$ plane in this work. The
instability region in the $v_0-T$ plane has been identified
numerically. Figure \ref{v0pb} compares the computed stability
boundary (blue) and that estimated from the spinodal boundary
(red). The discrepancy between the estimates and the numerical
results is a consequence of the shift in the spinodal region due
to the flow-induced density gradient and is evident from
Eq.~(\ref{eos}).

\begin{figure}
\centerline{\epsfig{file=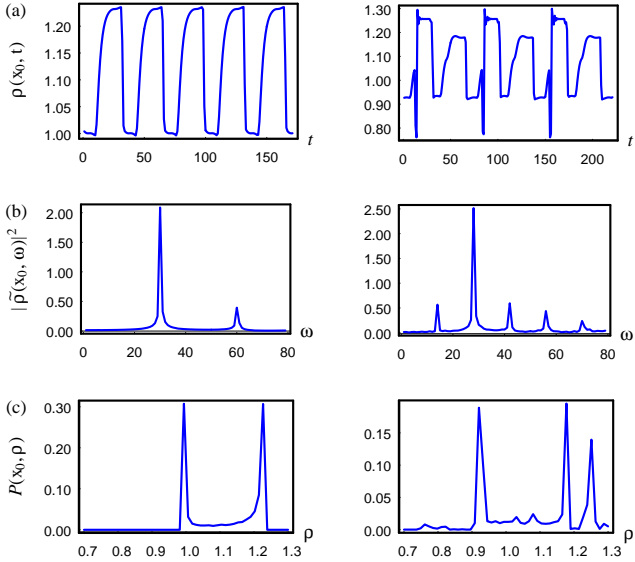, width=3.5in}} \caption{ Plots
of (a) $\rho(L,\;t)$, (b) $|\tilde{\rho}(L,\;\omega)|^2$ and (c)
$P(L,\;\rho)$ for the case of simple oscillation at $v_0$=0.142 on
the left panel, and for the case of complex oscillations at
$v_0=0.136$ on the right. The other parameters are set to
$v_L=0.1$, $\rho_0=0.8$, $T=0.987$, $\eta=0.01$, and
$\kappa=0.0001$.} \label{oscl}
\end{figure}

Our results further show that the instability leads to oscillatory
behavior in $\rho(L,\;t)$, the density at the outlet. The
density-oscillations have been classified into two classes: a
`simple' oscillation in which $\rho(L,\;t)$ oscillates between two
densities, and a  more `complex' oscillation where $\rho(L,\;t)$
cycles among three or more densities. The left panels of
Fig.~\ref{oscl} illustrate the behavior of a simple
density-oscillation whereas the right panels show a case of
complex oscillations. Figure \ref{oscl}(a) displays plots of
$\rho(L,\;t)$, showing the time evolution of the density at the
outlet. The plots of spectral density
$|\tilde{\rho}(L,\;\omega)|^2$ are shown in Fig.~\ref{oscl}(b). A
primary peak accompanied by its weaker harmonics is shown on the
left panel; a more complicated structure exists on the right
panel. Furthermore, the number of densities among which
$\rho(L,\;t)$ cycles can be picked out by evaluating the density
distribution function $P(L,\;\rho)\equiv {\cal
N}^{-1}\int\delta[\rho(L,\;t)-\rho]dt$ where ${\cal N}\equiv \int
P(L,\;\rho)d\rho$ is the normalization constant.  Figure
\ref{oscl}(c) displays $P(L,\;\rho)$ with two peaks on the left
 and three on the right.  The green curve in Fig.~\ref{v0pb} is the
boundary in the
$v_0-T$ plane separating the region of simple and complex
oscillations.  Simple oscillations occur in the region between the
blue and the green curves, while complex oscillations are found
under the green curve.

\begin{figure}
\centerline{\epsfig{file=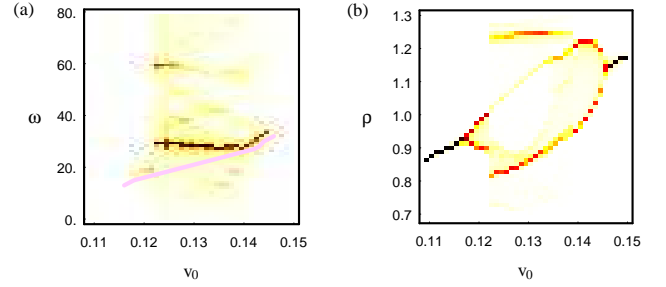, width=3.5in}} \caption{Density
plots of (a) $|\tilde{\rho}(L,\;\omega)|^2$ as a function of
$\omega$ and $v_0$, and (b) $P(L,\;\rho)$ as a function of $\rho$
and $v_0$. The pink curve in (a) is a guide-to-eye for the trend
of the principal peak in the region of simple oscillations.
Deviation from this curve indicates the region of complex
oscillation $v_0\in(0.123, 0.139)$. This region can be clearly
identified in (b) by the onset of additional peaks. The parameters
used are $\rho_0=0.8$, $v_L=0.1$, $T=0.987$, $\eta=0.01$ and
$\kappa=0.0001$.} \label{dens}
\end{figure}

Figure \ref{dens}(a) is a density plot of
$|\tilde{\rho}(L,\;\omega)|^2$ versus $v_0$ and $\omega$, and
Fig.~\ref{dens}(b) is a density plot of $P(L,\;\rho)$ as a
function of $v_0$ and $\rho$, showing the dependence of these
functions on the inlet velocity $v_0$, while keeping other
parameters fixed. Higher intensity is represented by darker colors
in the density plots, highlighting the peaks of the functions. The
appearance of peaks at nonzero frequency in
$|\rho(L,\;\omega)|^2$, and the multi-peak structures in
$P(L,\;\rho)$ reflect the unstable region in $v_0$
[$v_0\in(0.119,0.146)$] at $T=0.987$. The region of complex
oscillation at $T=0.987$ with $v_0\in(0.123,0.139)$ can be
identified as the deviation from the simple structure of a single
principal peak in $|\tilde{\rho}(L,\;\omega)|^2$ [the trend of the
principal peak is shown by the pink curve in Fig.~\ref{dens}(a)],
and the onset of additional peaks in $P(L,\;\rho)$ other than the
two expected for simple oscillations.

The scenario described by Eqs.~(\ref{continuity}) and
(\ref{NavierStokes}) with boundary conditions given by
Eqs.~(\ref{vinlet})-(\ref{noutlet}) can be related to the flow of
a Langmuir monolayer (a single molecular layer of insoluble
surfactant molecules spread at the air/water interface) during a
Langmuir-Blodgett  deposition. In a Langmuir-Blodgett deposition,
a substrate is withdrawn from or dipped into the water
transferring the surfactant molecules onto the substrate at the
contact line. The surfactant molecules are removed at the
substrate withdrawal speed at the contact line. The boundary
conditions at the outlet, Eqs.~(\ref{voutlet}) and
(\ref{noutlet}), are explicitly realized at the contact line if
the transfer ratio is unity at all times. Although the
correspondence between the inlet boundary conditions and the
actual experiments is not completely clear, Eqs.~(\ref{vinlet})
and (\ref{ninlet}) are reasonable for an appropriately chosen $L$.
A sensible estimate of the inlet-outlet separation $L$ for the
Langmuir-Blodgett deposition would be the meniscus height at the
contact line, which is $\sim10^{-3}$m. Hence, the specific choice
of boundary conditions, Eqs.~(\ref{vinlet})-(\ref{noutlet}), can
be realistic. Despite the attempt to associate the proposed
scenario to the actual experiment, the model is far from a
complete description of the system.  Many details relevant to the
Langmuir monolayer such as long-range dipolar interactions between
the surfactant molecules, surfactant-water interactions
\cite{Bruinsma}, water flow near the contact line
\cite{Petrov,Yulii}, elasticity of the monolayer \cite{Douglas},
are not taken into account. Nevertheless, the key emphasis is to
point out  that a discontinuous phase transition has been shown to
lead to simple period density-oscillations which provide a
possible mechanism for the formation of macroscopic arrays of
mesoscopic structures.

In summary, we have demonstrated numerically that a
one-dimensional flow of a van der Waals fluid near its
discontinuous liquid-gas transition exhibits a hydrodynamic
instability leading to periodic, classified as simple and complex,
oscillatory behavior. We highlight the key point of our analysis:
the instability in this simple system arises because of the
inability to support a steady-state solution  when enforcing the
continuity equation in the vicinity of a discontinuous transition.
Hence, this oscillatory behavior is not restricted to van der
Waals fluids and could be observed in other kinds of systems (with
well prescribed equation of state) that possess similar features.
We have attempted to relate this one-dimensional model to
Langmuir-Blodgett deposition where the boundary conditions can be
physically realized. This simple model provides us with a route to
periodic oscillatory behavior that has been observed \cite{exp}.
We note that the region in the parameter space where simple
oscillations occur is limited.  This is  a possible explanation
for the difficulty in finding the periodic oscillations and their
sensitivity to experimental conditions, such as pressure and
temperature. Estimates for the period and the interval of each of
the phases, within the framework of this model, require a
knowledge of the continuation of the equation of state into the
metastable regime and the spinodals.

We thank Dr.~Hans Riegler, Dr.~Jack F. Douglas, Professor Boris
Malomed, Professor Rashmi Desai  and Professor Yulii Shikhmurzaev
for many interesting discussions. One of us (KL) acknowledges a
Director's Fellowship at Los Alamos National Laboratory. This work
was supported by the US Department of energy.

\end{document}